\documentclass[10pt, a4paper,english]{article} % document type and language
\usepackage{csquotes}

\usepackage{amsmath, amssymb, amsthm}
\usepackage{pictex, dcpic}

\def\al{\alpha}
\def\be{\beta}
\def\de{\delta}
\def\ga{\gamma}

\def\ep{\epsilon}
\def\io{\iota}

\def\la{\lambda}

\def\om{\omega}
\def\si{\sigma}
\def\vp{\varphi}

\def\ka{\kappa}

\def\Ga{\Gamma}

\def\La{\Lambda}

\def\Si{\Sigma}

 \def\calC{{\hbox{\cal C}}}

 \def\su{{\mathfrak{su}}}
 
 \def\gotg{{\mathfrak{g}}}
 \def\goth{{\mathfrak{h}}}
 \def\gotm{{\mathfrak{m}}}
 \def\gotk{{\mathfrak{k}}}
 \def\spin{{\mathfrak{spin}}}
 \def\slC{{\mathfrak{sl}}}

 \def\C{\mathbb{C}}

 \def\R{\mathbb{R}}

\def\Con{{\hbox{Con}}}
\def\Aut{{\hbox{Aut}}}

\def\Ad{{\hbox{Ad}}}

\def\Im{{\hbox{Im}}}

\def\Spin{{\hbox{Spin}}}
\def\SO{{\hbox{SO}}}
\def\SU{{\hbox{SU}}}
\def\SL{{\hbox{SL}}}
\def\GL{{\hbox{GL}}}
\def\det{{\hbox{det}}}

\def\Diff{{\hbox{Diff}}}
\def\di{{\hbox{d}}}

\def\diag{{\hbox{diag}}}

\def\ip{\hbox to4pt{\leaders\hrule height0.3pt\hfill}\vbox to8pt{\leaders\vrule width0.3pt\vfill}\kern 2pt}

\def\del{\partial}
\def\na{\nabla}

\def\arr{\rightarrow}

\def\harr{\hookrightarrow}
\def\hlarr{\lhook\joinrel\longrightarrow}
\def\then{\Rightarrow}

\def\barJ{\bar J}

\def\frac[#1/#2]{\hbox{$#1\over#2$}}
\def\Frac[#1/#2]{{#1\over#2}}
\def\({\left(}
\def\){\right)}
\def\[{\left[}
\def\]{\right]}
\def\^#1{{}^{#1}_{\>\cdot}}
\def\_#1{{}_{#1}^{\>\cdot}}
\def\Label=#1{{\buildrel {\hbox{\fiveSerif \ShowLabel{#1}}}\over =}}
\def\<{\kern -1pt}

\def\ShowLabel#1{\ref{#1}}

\def\ms{\medskip}
\def\ss{\smallskip}
\def\ni{\noindent}

\def\eq#1{\begin{equation}#1\end{equation}}
\def\eqLabel#1#2{\begin{equation}#1\label{#2}\end{equation}}

\def\Cases#1{\begin{cases}#1\end{cases}}
\def\Matrix#1{\begin{matrix}#1\end{matrix}}
\def\Align#1{\begin{aligned}#1\end{aligned}}

\def\eqs#1{\eq{\Align{#1}}}

\long\def\Note#1{\blockquote{\footnotesize #1}}

\date{}

\def\Figure[#1]#2{\begin{figure}[htbp] %  figure placement: here, top, bottom, or page
   \centering
   \includegraphics[#1]{#2} }

\def\EndFigure{\end{figure}}

\def\Diagram#1{\eq{
\begindc{\commdiag}[10]
#1
\enddc%
}%
}

\def\Itemize#1{\begin{itemize}#1\end{itemize}}
\def\Item[#1]{\item[#1]}

\title{Introduction to Loop Quantum Gravity.\\
The Holst's action and the covariant formalism}

\author{\small L.Fatibene$^{a,b}$, A.Orizzonte$^{a}$, \\
\small A.Albano$^a$, S.Coriasco$^{a}$, M.Ferraris$^a$, S.Garruto$^c$, N.Morandi$^d$\\
\\
\small$^a$ Department of Mathematics, University of Torino (Italy)\\
\small$^b$ Ist. Naz. Fisica Nucleare (INFN) - Sezione Torino - Iniziativa spec. QGSKY (Italy)\\
\small$^c$ Department of Business and Management,  LUISS Guido Carli, Roma (Italy)\\
\small$^d$  Department of Econometrics and Operations Research,  Tilburg University (Netherlands)
}

\begin{document}

%%%%%%%%%%%%%%%%%%%%%%%%%%%%%%%%%%%%%%%%%%%%%%%%%%%

\maketitle

\begin{abstract}
We review Holst formalism and we discuss dynamical equivalence with standard GR (in dimension $4$).
Holst formalism is written for a spin coframe field $e^I_\mu$ and a $\Spin(3,1)$-connection $\om^{IJ}_\mu$ on spacetime $M$ and it depends 
on the {\it Holst parameter} $\ga\in \R-\{0\}$.

We show the model is dynamically equivalent to standard GR, in the sense that up to a pointwise $\Spin(3,1)$-gauge transformation acting on  (uppercase Latin) frame indices, 
solutions of the two models are in one-to-one correspondence.
Hence the two models are classically equivalent.

One can also introduce new variables by splitting the spin connection into a pair of a $\Spin(3)$-connection $A^i_\mu$ and a $\Spin(3)$-valued 1-form $k^i_\mu$.
The construction of these new variables relies on a particular algebraic structure, called a {\it reductive splitting}.
A reductive splitting is a weaker structure than requiring that the gauge group splits as the products of two sub-groups, as it happens in Euclidean signature in the selfdual formulation 
originally introduced in this context by Ashtekar, and it still allows to deal with the Lorentzian signature without resorting to complexifications.

The reductive splitting of $\SL(2, \C)$ is not unique and it is parameterized by a real parameter $\be$ which is called the {\it Immirzi parameter}.
The splitting is here done {\it on spacetime}, not on space as it is usually done in the literature, to obtain a $\Spin(3)$-connection $A^i_\mu$, which is called the {\it Barbero--Immirzi connection} on spacetime.
One obtains a covariant model depending on the fields $(e^I_\mu, A^i_\mu, k^i_\mu)$ which is again dynamically equivalent to standard GR (as well as the Holst action).

Usually in the literature one sets $\be=\ga$ for the sake of simplicity. 
Here we keep the Holst and Immirzi parameters distinct to show that eventually only $\be$ will survive in boundary field equations. 
\end{abstract}

\section{Foreword}

This is the first paper in a series of lecture notes aiming to provide a coherent and homogeneous introduction to Loop Quantum Gravity (LQG).
LQG has grown considerably in the last decades and it now includes much more material than what one can hope to include in a relatively general lecture note series and there are books to provide physically well motivated accounts of the theory;
see \cite{Ash}, \cite{Rovelli1}, \cite{Rovelli2}.
The aim of this project is not to cover all material which is now considered relevant to LQG but to select a coherent and notationally homogeneous path to the core of the theory  and to  cover it
in enough detail to be followed by researchers and students who would like to get into the field.
More will be needed, something will have to be forgotten, but we believe this will give a solid basis to build upon.
We do not stress much on the physical motivations, we focus on mathematical aspects and structure of the theory which we think will turn out to be good tools when physics motivations will have to be discussed.
Anyway, each lecture is meant to be self-contained, focused on a particular aspect of the theory.

Some of the topics along the way are not covered in the standard way. 
For example, we start by defining Barbero-Immirzi $\SU(2)$-connection {\it on spacetime} while usually this is defined on a spatial leaf of an ADM foliation.
We shall show this is possible (contrary to what sometimes argued; see \cite{Samuel}, \cite{HolonomyClassified}) and it gives a better view of how the classical theory is defined and adapted to the process one wishes to later apply to ``guess'' the quantum theory.

Another example is when later on we shall regard spin networks as encoding functionals of the connection (as it is traditional, see e.g.~\cite{Rovelli1}, though not too systematically discussed). 
We will practice a bit about this correspondence so that graphic methods do not take over.
This will allow us to better discuss operators, and to render clearly the bridge between operators on spin networks and discrete geometries which will be the starting point of covariant formulation of spin foams, see \cite{Rovelli2}.
 
We shall also focus on globality of the mathematical structures in order to discuss the relation between global mathematical structures and physical motivations.
For example, we shall systematically discuss connections, which are global connections on principal bundles as done in geometry; see \cite{Steenrod}, \cite{KN}, \cite{book1}, \cite{book2}.
We shall discuss how this relates with physical motivation. 
The issue is still open, many researchers think a local language is simpler and sufficient, we shall argue that the two viewpoints are eventually equivalent and discussing the issue explicitly is useful whatever viewpoints one ends up with.
Moreover, by pinpointing that global properties are hidden in transformations laws, one has a structure to adjust (a sort of) relational viewpoint already in a classical framework. 
In fact, we shall argue intrinsic properties and transformation laws are what really encode the physical knowledge (in view of the relativity principle) and they are written in the relations among observers, which are identified with conventions for describing physical observations in terms of numbers, namely, in a relativistic context, with coordinate charts on spacetime.

We think that global notation is not much harder to develop, it is clearer, and when one decides to work locally, as a matter of fact, as soon as transformation laws of objects
 are taken into account,  the global properties are recovered and the whole issue becomes only a matter of notation. Considering transformation laws is equivalent to global viewpoint even working with local representatives of objects in coordinates.
And GR without transformations laws is not GR.
To the very least this discussion will allow us to clarify what {\it background free} means, see \cite{Pathway}, \cite{Giulini}, why physics should care, and why different attitudes are reasonable in different situations.
This is actually an important issue even though rather non-technical, hence we discuss it in the Appendix A.

\section{Introduction}

Loop Quantum Gravity (LQG) is a (if you want, proposal for a)  background free quantization of the gravitational field as described by standard General Relativity (GR) in dimension 4.

We discuss the Holst model  (see \cite{HolstOriginal}, \cite{NostroRov1}, \cite{NostroRov2}, \cite{NostroHolst}) which is classically equivalent to standard General Relativity (GR), as well as the starting point of quantization {\it \`a la} LQG.
Here we shall focus on the classical setting and equivalence with standard GR.

Starting from the Holst model, we will make a field transformation and define the Barbero--Immirzi formulation, which is similar to what is done in LQG
before starting quantization. However, we define the Barbero-Immirzi connection {\it on spacetime}, while usually it is defined on a leaf of an ADM foliation, see \cite{ADM}, \cite{NostroADM}, \cite{Thiemann2}.
We shall see that  starting from a covariant model will clarify the constraint structure of the theory and allow us to {\it derive} some of the relations which are {\it assumed} as definitions
in the usual spatial formulation.
We shall also clarify the role of parameters appearing in defining the  Holst action and the Barbero--Immirzi connection.
In both cases, our aim is not to (and we do not) obtain new results, we simply confirm the choices which are usually taken by definition by showing that one has not really other options.
\ms

There are many different,  although somehow equivalent, formulations of standard GR in dimension 4, based on different representations of the gravitational field 
in terms of  various geometric objects.
These are {\it dynamically equivalent} field theories. 
Strictly speaking, by {\it dynamically equivalent theories} one means that there is a one-to-one correspondence between solutions of the two theories.

However, in different formulations one often has different gauge groups. 
In standard GR one considers a Lorentzian metric $g$ as a fundamental field, and it is based on the Hilbert (second order) Lagrangian. 
Accordingly, the gravitational field is identified with classes of metrics up to diffeomorphisms.
In other formulations, for example frame formulations, one uses a frame $e_I$ as a fundamental field, which is defined up to an automorphism of a structure bundle $P$, 
which in the Holst case means that besides diffeomorphisms, frames are defined up to a pointwise Lorentz (or $\Spin(3,1)$) transformation.

When one compares a model based on metrics and a model based on a frames, of course there are infinitely many frames associated to the same metric due to the pointwise Lorentz transformations.
However, one still has a one-to-one correspondence between classes of metrics up to diffeomorphisms and classes of frames up to diffeomorpfisms and Lorentz transformations.
Let us still say the two models are dynamically equivalent in this case, thus we define {\it dynamical equivalence} when one has a one-to-one correspondence between {\it states}
of the gravitational field, in one theory and the other.
In this way, the standard metric and frame formulations are equivalent (as we shall discuss here {\it in vacuum}).

Accordingly, a model of gravity is dynamically equivalent to standard GR if it induces from its solutions the same metrics standard GR does.
Often different formulations of gravitational theories are classified in terms of the fundamental fields they use. 
Fundamental fields are important since their choice dictates the way we deform the action, e.g.~to derive field equations.

Hence, we have {\it purely metric} models in which there is only a Lorentzian metric $g$, 
{\it metric-affine} (or {\it Palatini}) formulations when one has a metric $g$ and an independent connection $\tilde \Ga$ (with or without torsion),
or {\it purely affine} formulations when the action depends only on a connection $\tilde \Ga$ (with or without torsion).

Then we can use frames (to be defined precisely below) instead of the metric and we have {\it purely frame} and {\it frame-affine} formulations.
The Holst formulation is based on a special case of frame $e_I$, called a {\it spin frame} and an independent \Spin(3,1)-connection $\om^{IJ}$, hence it is a frame-affine formulation; see \cite{KN}, \cite{book1}, \cite{book2}, \cite{Lawson}, \cite{Jadwisin},
\cite{Spinors}, \cite{Noris}.
Of course, frame and frame-affine formulations are preferable over metric and metric-affine ones, since sooner or later one will wish to describe spinors and, as we shall argue, spin frames are {\it also}
the exact structure needed to deal with (global) Dirac equations in interaction with the gravitational field; see \cite{Lawson}, \cite{Spinors}.

In general, spin frames exist on a manifold if some topological restrictions on the manifold are satisfied; see \cite{Orizzonte}. 
These restrictions are {\it global} restrictions on the spacetime manifold $M$ which are encoded into a characteristic class and they have to be satisfied if one eventually wants to have global Dirac equations for spinors.
When spin frames exist, they {\it also} define an associated metric, which in this case appears as a by product of the spin frame and, as such,  is not a fundamental field in the model.  

\section{Holst formulation}

The {\it Holst formulation} is a field theory (see \cite{HolstOriginal}, \cite{NostroRov2}, \cite{NostroHolst}) defined for a spin frame $e_I^\mu$ (or, equivalently, the spin coframe $e^I_\mu$) and a $\Spin(3,1)$-connection $\om^{IJ}_\mu$ (on a {\it structure bundle} $P$) depending on a real parameter $\ga\in \R-\{0\}$, called the {\it Holst parameter},
which is eventually dynamically equivalent to standard GR.

Let us start by defining a {\it spin frame}; see \cite{Spinors}, \cite{Jadwisin}.
Consider a manifold $M$ and its frame bundle $L(M)$, which is a $\GL(m)$-principal bundle. If we choose coordinates $x^\mu$ on $M$ that induces {\it natural coordinates} $(x^\mu, e_I^\mu)$ on $L(M)$, where $|e_I^\mu|\in \GL(m)$.
We fix a signature $\eta=(3,1)$, the relative spin group $\Spin(3,1)\simeq \SL(2, \C)$, and a {\it structure bundle} $P$ which is a $\SL(2, \C)$-principal bundle $[\pi \colon P\arr M]$; \cite{book1}, \cite{Kolar}.
As on any principal bundle, on $P$ one has transition functions between local trivializations $t\colon \pi^{-1}(U)\arr U\times \SL(2, \C)\colon p\mapsto (x, S)$ in the form
\eq{
\Cases{
 x'=x \cr
 S' = \vp(x)\cdot S
}
}
for some local function $\vp \colon U\arr \SL(2, \C)$, which is called the {\it transition function} (in the group).

On the structure bundle, we have coordinates $(x^\mu, S)$ with $S\in \SL(2, \C)$ and the canonical right action $R_g\colon P\arr P \colon p\mapsto p\cdot g$, which is well defined (independent of the trivialization) and a vertical, transitive on the fibers, and free action.

\Note{
In view of the canonical right action, one has a one-to-one correspondence between local trivializations and local sections. Given a local section $\si \colon U\arr \pi^{-1}(U)\subset P$ and $p\in \pi^{-1}(U)$ we set $x=\pi(p)$ and
we can define a local trivialization  $t \colon \pi^{-1}(U)\arr U\times \SL(2, \C) \colon p\mapsto (x, S)$ where $p = \si(x)\cdot S$, which uniquely determines $S$ since the right action is free.
}

Then we define a {\it spin frame} to be a (global) map $e \colon P\arr L(M)$ which is {\it vertical} and {\it equivariant} with respect to the group homomorphism
$ \ell \colon \Spin(3, 1) \arr \SO(3, 1) \harr \GL(4)$, i.e.~it preserves the right action as
\Diagram{
\obj(20,50)[P]{$P$}
\obj(70,50)[LM]{$L(M)$}
\obj(70,10)[M2]{$M$}
\obj(20,10)[M1]{$M$}
\obj(180,40)[eqivariance]{$e(p\cdot S) = e(p) \cdot  \ell(S)$}
\mor{LM}{M2}{}
\mor{P}{M1}{$\pi$}
\mor{P}{LM}{$e$}[\atleft,\solidarrow]
\mor{M1}{M2}{}[\atright,\solidline]
\mor(20,12)(70,12){}[\atright,\solidline]
}
\Note{
As a matter of fact, spin frames do not always exist for a generic choice of $M$ and $P$ over it.
Since $[M\times \SL(2, \C)\arr M]$  is a (trivial) principal bundle, a structure bundle always exists but not every structure bundle $P$ admits a global spin frame.
For example if $L(M)$ is not trivial (i.e.~$M$ is not parallelizable) and $P$ is trivial one can show immediately that there is no global spin frame $e \colon P\arr L(M)$.
If it existed, a spin frame would be represented locally by a matrix $e_I^\mu(x) \in \GL(4)$ such that
\eqLabel{
e(\si(x)) = e_I^\mu(x) \del_\mu
}{LocalSpinFrame}
and we denote by $e^I_\mu$ its inverse matrix.
But if the change the local trivialization  on $P$ (i.e.~the section $\si$ to $\si'(x)= \si(x)\cdot \bar \vp(x)$) and coordinates $x'^\mu= x'^\mu(x)$ on $M$, transformation laws are
\eqLabel{
e'^\mu_I = J^\mu_\nu e_J^\nu  \ell^J_I(\bar \vp)
}{TransformationLawsSPinFrames}
where the bar denotes the inverse matrix or group element and $J^\mu_\nu$ is the Jacobian of the change of coordinates, i.e.~transition functions on $M$. 
Together, the local representation (\ShowLabel{LocalSpinFrame}) and the transformation law (\ShowLabel{TransformationLawsSPinFrames}),
are equivalent to an intrinsic and global description of the spin frame $e \colon P\arr L(M)$.

 Let us stress that a spin frame is not a section of $L(M)$. 
It can rather be seen as a family of local sections of  $L(M)$ which differ on the overlaps by an $\SL(2,\C)$ transformation in the representation $\ell$.
Again, this is not an issue about globality, it is an issue about transformation laws (and eventually about equivalence classes representing physical states).
If they were to be considered as (local) sections of $L(M)$, i.e.~as frames, they would transform as sections of $L(M)$, namely as $e'^\mu_I = J^\mu_\nu e^\nu_I$, with no $\SL(2,\C)$ transformation,
which eventually would lead to a different covariant derivative. The fact that in the physics literature the covariant derivative prescribed for spin frames is used is a clear indication that they are in fact using spin frames
even if, working locally, they are considered as (local) sections of $L(M)$ (with the side effect that one has to reintroduce $\SL(2, \C)$ transformations by an {\it ad hoc} procedure).

For a spin frame $e \colon P\arr L(M)$ the image  $\Im(e)= \SO(M,g)\subset L(M)$ is the sub-bundle of $g$-orthonormal frames on $M$, where the {\it metric $g$ induced by the spin frame} is
\eq{
g= (e^I_\mu \eta_{IJ} e^J_\nu)\> dx^\mu\otimes dx^\nu
}
Then the pair $(P, \hat e)$ with $\hat e \colon P\arr \SO(M, g)$ is a standard {\it spin structure} on $(M, g)$. 
For a standard  spin structure to exist, one needs global Lorentzian metrics to exist on $M$, i.e.~the tangent bundle $TM$ splits as the sum of a {\it time bundle} $T_1$ and a {\it space bundle} $T_3$, of rank $1$ and $3$ respectively.
Moreover, one needs the second Stiefel-Whitney class of $M$ to be vanishing; see \cite{Lawson}, \cite{Orizzonte}, \cite{KN}. 
If this second condition is met, then one knows there exists some structure bundle $P$ which allows global spin frames $e \colon P\arr L(M)$, even though sometimes not all principal bundles on $M$ are allowed as structure bundles, e.g.~sometimes one has principal bundles that do not allow global spin frames anyway.
A manifold $M$ satisfying these conditions is called a {\it spin manifold} 
%(or we can say that {\it the manifold $M$ is spin}) 
and we just argued that spacetimes need to be spin manifolds. 
Given a spin manifold $M$, one can find a structure bundle $P$ over it and a global spin frame $e \colon P\arr L(M)$ on it,  possibly having more than one choice available as a structure bundle.
These topological constraints are there only as long as one requires the objects to be global since, locally, one always has spin frames.

Finally, let us mention that given a structure bundle $P$ one can functorially define a {\it spin frame bundle} $F(P)$ whose global sections are in one-to-one correspondence with global spin frames.
Local coordinates on $F(P)$ are of course $(x^\mu, e_I^\mu)$ which transforms according to (\ShowLabel{TransformationLawsSPinFrames}).
}

Once we select a structure bundle $[\pi \colon P\arr M]$ we can also define principal connections on $P$.

\Note{
On the structure bundle $P$, one can locally fix a right invariant pointwise basis $\si_{IJ}$ (skew in $[IJ]$) of vertical vectors one for each basis %$T_{IJ}$ 
in the Lie algebra $\slC(2, \C)$.
Then, a principal connection is
\eq{
\om = dx^\mu\otimes (\del_\mu - \om^{IJ}_\mu(x) \si_{IJ})
}
which in fact is map $\om \colon \pi^{\ast} TM \arr TP$. The connection $\om$ induces, at each point $p\in P$, linear maps $\om_p \colon T_xM \arr T_pP$ which define the spaces of horizontal vectors $H_p= \om_p(T_xM)\subset T_pP$
and the horizontal lift(s) $\om_p(v)= v^\mu (\del_\mu - \om^{IJ}_\mu(x) \si_{IJ})$ to $T_pP$ of a tangent vector $v\in T_xM$.

By changing coordinates on $M$ and local trivializations on the structure bundle $P$, one gets transformation laws for the local representations of a connection by its coefficients as
\eqLabel{
\om'^{IJ}_\mu = \barJ_\mu^\nu  \ell^I_K(\vp)  \(\ell^J_L(\vp) \om^{KL}_\nu +  \del_\nu \ell^{K}_L(\bar\vp) \eta^{LJ}\)
}{TrasformationLawsConnection}
where $\ell:\SL(2,\C)\arr \SO(3,1)$ is the {\it covering map} (to be discussed later on when we discuss spin groups more generally in their Clifford Algebras; see \cite{LN6}).

As for spin frames, the local expression $\om^{IJ}_\mu$ together with the transformation laws (\ShowLabel{TrasformationLawsConnection}) are equivalent to 
a global and intrinsic description of a connection.
As in the case of spin frames, one can functorially define a bundle $\Con(P)$ with coordinates $(x^\mu, \om^{IJ}_\mu)$ so that there is a one-to-one correspondence between global sections of $\Con(P)$ and global connections on $P$.
}

Thus we have our fundamental fields $(e_I^\mu, \om^{IJ}_\mu)$, they can be even regarded as global sections in a suitable configuration bundle $F(P)\times_M \Con(P)$ if needed, 
although it is irrelevant here.
Now we need a dynamics given by a Lagrangian.
We decide $e_I^\mu$ enters at order zero (no derivatives), while the spin connection $\om^{IJ}_\mu$ enters at order 1 (i.e.~with its first derivative).
Moreover, the action must be covariant with respect to transformation laws (\ShowLabel{TransformationLawsSPinFrames}) and (\ShowLabel{TrasformationLawsConnection}) combined (which are an action of the group $\Aut(P)$ of automorphisms of the structure bundle, acting on the configuration bundle, but again here we ignore it).

At first order of the connection $\om^{IJ}_\mu$, we can define the curvature
\eq{
R^{IJ}{}_{\mu\nu}= \del_\mu \om^{IJ}_\nu -\del_\nu \om^{IJ}_\mu 
 + \om^I{}\_{K\mu} \om^{KJ}_\nu - \om^I{}\_{K\nu} \om^{KJ}_\mu
}
where uppercase Latin indices are moved consistently by the matrix $\eta_{IJ} = \diag(-1, 1, 1, 1)$. Notice that $R^{IJ}{}_{\mu\nu}$ is a tensor, i.e.~it transforms as
 \eq{
 R'^{IJ}{}_{\mu\nu}=  \ell^I_K(\vp) \ell^J_L(\vp) R^{KL}{}_{\al\be}  \barJ_\mu^\al\barJ_\nu^\be
 }

\Note{
 One can obtain the Ricci and scalar curvature by contraction, namely
 \eq{
R^I_\mu = e_J^\nu R^{IJ}{}_{\mu\nu}
\qquad\qquad
R = e_I^\mu R^I_\mu
}
which of course depend on the connection and the spin frame.
}

We can define a curvature 2-form as well as a coframe 1-form  (valued in the algebra $\slC(2, \C)$)
\eq{
R^{IJ} = \frac[1/2] R^{IJ}{}_{\mu\nu} dx^\mu \land dx^\mu
\qquad\qquad
e^I = e^I_\mu dx^\mu
}

In dimension 4, the {\it Holst action} then is
\eq{
A_D [e, j^1\om] =\frac[1/4\ka] \int_D R^{IJ}\land e^K\land e^L \ep_{IJKL} +\frac[2/\ga] R^{IJ}\land e\_I\land e\_J - \frac[\La/6]  e^I\land e^J \land e^K\land e^L \ep_{IJKL} 
}
The real (adimensional) parameter $\ga\in \R-\{0\}$ is called the {\it Holst parameter}, $\La$ is the {\it cosmological constant}, and $\ka= 8\pi G c^{-3}$.
 For later convenience, let us define the 2-form
\eq{
B_{IJ}= \frac[1/2]  e^K\land e^L \ep_{IJKL} +\frac[1/\ga] e\_I\land e\_J 
}
so that we can re-write the action as
\eq{
A_D [e, j^1\om] =\frac[1/2\ka] \int_D R^{IJ}\land B_{IJ} - \frac[\La/12]  e^I\land e^J \land e^K\land e^L \ep_{IJKL} 
}

\Note{
If $B_{IJ}$ were a fundamental field, this theory (with $\La=0$) would be a {\it BF-theory}. That is well known and studied. Field equations would be
\eq{
R^{IJ}=0
\qquad
\hat\na_\mu B_{IJ} =0
}
since the variation of the Lagrangian would be
\eq{
\de L =   - \hat\na B_{IJ}\land \de \om^{IJ} + R^{IJ}\land \de B_{IJ} +\hat\na \( B_{IJ}\land \de \om^{IJ}\)
}
where $\hat \na$ denotes the covariant derivative induced by $\om^{IJ}$ (and the Levi-Civita connection $\{g\}$ of the induced metric $g$ for world indices).
However, in our case $B_{IJ}$ is not a fundamental field, it is a function of the spin (co)frame, which is fundamental instead.
A fundamental 2-form $B_{IJ}$ has 36 independent components and variations, a spin frame only 16.
That means that for us the allowed deformations of $B_{IJ}$ are only the 16 dictated by the functional form of $B_{IJ}$, namely
\eq{
\de B_{IJ} = \( \ep_{IJKL}  e^K  +  \frac[2/\ga]e_{[I} \eta_{J]L}\)\land \de e^L
}
}

Thus the field equations of the Holst model are
\eqLabel{
\Cases{
 \ep_{IJKL}   \hat \na e^K\land e^L  + \frac[2/\ga]  \hat \na e_{[I}\land e_{J]} =0			\cr
R^{IJ} \land \( \ep_{IJKL}  e^K  +  \frac[2/\ga]  e_{[I} \eta_{J]L}\)  =  \frac[\La/3]  \ep_{IJKL} e^I\land e^J\land e^K 
 }
 }{HolstFieldEquations}

\

\section{Dynamical equivalence with standard GR}

We have to show that, although field equations (\ShowLabel{HolstFieldEquations}) depend on an extra parameter $\ga$, they are all dynamically equivalent to standard GR, i.e.~they
define the same set of Lorentzian metrics as solutions.

The first field equation is actually algebraic and linear in the connection $\om^{IJ}$. It is not a surprise we can solve it explicitly.

\Note{
The manipulation is quite complicated, though elementary.
We first need to know that the spin frame induces a connection $\tilde \Ga$ on $P$ such that $\tilde \na_\mu e^I_\nu =0$.
One readily has 
\eq{
\tilde \na_\mu e^I_\nu =0
\iff
\tilde \Ga^{IJ}_\mu= e^I_\al \( \{g\}^\al_{\be\mu} e^\be_K + d_\mu e^\al_K   \)\eta^{KJ}
}
which is called the {\it spin connection}.
Since it is a connection on $P$ as $\om^{IJ}$ is, then their difference is a tensor, thus let us set
\eq{
z^{IJK} := (\om^{IJ}_\mu - \tilde \Ga^{IJ}_\mu)e\^{K\mu}
\qquad\then z^{(IJ)K}=0
}
Let us stress that we need both $\om$ and $\tilde \Ga$ to be global connections, so that their difference is a tensor.
We shall show that $z=0$, which makes sense intrinsically just because it is tensor, therefore we are implicitly using the  transformation laws in the proof. One cannot ignore global properties. More examples will follow.

Now we can go back to the field equations and use the identity $\hat \na e^I = \tilde \na e^I + z^I{}_K\land e^K$ to transform it into an algebraic linear equation for $z$, namely
\eqs{
&\( \ga^2   +1\) \hat \na e^{[I}\land e^{J]} =0
\quad\then 
\hat \na e^{[I}\land e^{J]} =0
\quad\then\cr
\then\quad&
z^{[I}{}_{K\nu} e^K_\rho e^{J]}_{\si} \ep^{\mu\nu\rho\si}=0
\quad\then
z^{I}{}_{K[\nu}  e^{J}_{\si} e^K_{\rho]} =z^{J}{}_{K[\nu}  e^{I}_{\si} e^K_{\rho]}  \cr
}
By tracing this identity we obtain $z^J{}_{IJ}  =  0$ and plugging it back into the field equations we can rewrite the first field equation as
\eq{
 z^{I[JK]}    = 0
} 
Now that we know that $ z^{I[JK]}    = 0$ and $ z^{(IJ)K}=0$, we can apply a standard argument to show that $ z^{IJK}=0$, i.e.
\eq{
z^{IJK} = z^{IKJ} = - z^{KIJ} = -z^{KJI} = z^{JKI} = z^{JIK} = - z^{IJK}
\quad\then 
z^{IJK}=0
}
}

Therefore we have $\om^{IJ}=\tilde \Ga^{IJ}$ along solutions from the first field equation.
Now we can check that the curvature $\tilde R^{IJ} $ of the connection $\tilde \Ga$ can be written in terms of the Riemann tensor $\tilde R^\al{}_{\be\mu\nu}$ of the induced metric $g$ as
\eq{
\tilde R^{IJ} = \frac[1/2]e^I_\al e^{J\be} \tilde R^\al{}_{\be\mu\nu} dx^\mu\land dx^\nu
}
and substitute that into the second field equation. This way we obtain
\eq{
-4 \ga     e   \(\tilde  R^{\si}{}_{\mu} -\frac[1/2] \tilde  R \de^\si_\mu    \)  e_K^\mu   = 4e\ga \La   \de^\si_\mu e_K^\mu	
\quad\then  \tilde R_{\mu\nu} -\frac[1/2]\tilde  R g_{\mu\nu}= -\La g_{\mu\nu}
}
where we used the first Bianchi identity $\tilde R^\al{}_{[\be\mu\nu]}=0$ and we set $e=\det(e^I_\mu)$. Here is where we need $\ga$ to be non-zero.

This last equation is purely metric and it singles out as solutions the same metrics as standard GR.
Accordingly, the Holst formulation and standard GR are dynamically equivalent.

\section{Barbero-Immirzi formulation}

Now we want to write the Holst formulation in terms of new fields; see \cite{Ash}, \cite{NostroAsh}, \cite{BI1}.
Once again, we need to fix some topological argument first, in order to keep global properties under control.
Since this is simply an (algebraic) field transformation, we shall not even need to prove dynamical equivalence since it will follow directly from the fact that the Lagrangian is global.

We discussed the Holst formulation which is written for fields defined on the structure bundle $P$, which is a $\SL(2, \C)$-principal bundle.
As a matter of fact, we have a closed subgroup $\Spin(3,0)\simeq \SU(2)\subset \SL(2, \C)$ and we can ask whether we can restrict trivializations on $P$ so that transition functions are valued into $\SU(2)\subset \SL(2, \C)$.

\Note{
We have similar examples on the frame bundle $L(M)$.
Initially, $L(M)$ is a principal bundle with the  group $\GL(4)$ and one defines a local trivialization for any local frame. 
For example, for natural frames $\del_\mu$ transition functions $J^\mu_\nu$ are clearly in $\GL(4)$.

However, if there is a strictly Riemannian metric $h$ defined on $M$ one can always use $h$-orthonormal frames $V_a$ to define local trivializations on $L(M)$ and, in that case, transition functions are clearly in the orthogonal group $O(4)\subset \GL(4)$.
As a matter of fact, this is always possible, since one can show that on any manifold one can define a strictly Riemannian metric $h$.

If the manifold $M$ is orientable, one can define {\it positively oriented frames} and further reduce the structure group to $\SO(4)\subset O(4)\subset \GL(4)$.
Of course, the first reduction to $O(4)$ is always possible, the second reduction to $\SO(4)$ needs a topological condition (orientability) to be met.

Also, if we want to reduce to the orthogonal group in a different signature, we need topological conditions.
That is because we need topological conditions for the metric in non-Euclidean signature to exist.
Once it does exist, orthonormal frames do exist and they define reductions as in the strictly Riemannian case.
}

In the physical language, we have a gauge theory for the group $\SL(2, \C)$ and we want to discuss whether we are able to partially gauge fix to a subgroup $\SU(2)$.
This can be done if and only if we can find an $\SU(2)$-principal bundle $[\tau:\Si\arr M]$ and a bundle map $\io\colon\Si\arr P$ such that
%\Diagram{
%\obj(50,50)[beP]{$\Si$}
%\obj(100,50)[P]{$P$}
%%\obj(150,50)[LM]{$L(M)$}
%\obj(50,10)[M0]{$M$}
%\obj(100,10)[M1]{$M$}
%%\obj(150,10)[M2]{$M$}
%%
%%\obj(180,40)[eqivariance]{$e(p\cdot S) = e(p) \cdot \hat \ell(S)$}
%%
%%\mor{LM}{M2}{$\>_{GL(m)}$}
%\mor{P}{M1}{$\>_{SL(2,C)}$}
%\mor{beP}{M0}{$\>_{SU(2)}$}
%%
%\mor{beP}{P}{$\io$}[\atleft,\solidarrow]
%%\mor{P}{LM}{$e$}[\atleft,\solidarrow]
%%
%%\mor{M1}{M2}{}[\atright,\solidline]
%%\mor(100,13)(150,13){}[\atright,\solidline]
%\mor{M0}{M1}{}[\atright,\solidline]
%\mor(50,13)(100,13){}[\atright,\solidline]
%}
%and the bundle map $\io:\Si\arr P$ 
that is vertical and equivariant with respect to the group homomorphism $i \colon \SU(2)\arr \SL(2, \C)$, namely we have
\Diagram{
\obj(20,50)[P]{$\Si$}
\obj(70,50)[LM]{$P$}
\obj(70,10)[M2]{$M$}
\obj(20,10)[M1]{$M$}
\obj(180,40)[eqivariance]{$\io(p\cdot U) = e(p) \cdot  i(U)$}
\mor{LM}{M2}{$\pi$}
\mor{P}{M1}{$\tau$}
\mor{P}{LM}{$\io$}[\atleft,\solidarrow]
\mor{M1}{M2}{}[\atright,\solidline]
\mor(20,12)(70,12){}[\atright,\solidline]
}
The pair $(\Si, \io)$ is called a {\it reduction} to the subgroup $\SU(2)$. 

\Note{
As we discussed, in general, existence of reductions depends on topological conditions which only occasionally are automatically satisfied.
Well, one can prove this is a case in which the reduction comes for free, it always exists a reduction to the group $\SU(2)$.

Actually this generalizes to any spin manifold $M$ of dimension $n+1$, where one has a reduction from $\Spin(n, 1)$ to $\Spin(n, 0)$, for free.
}

Then we are in the situation where 
\Diagram{
\obj(50,50)[beP]{$\Si$}
\obj(100,50)[P]{$P$}
\obj(150,50)[LM]{$L(M)$}
\obj(50,10)[M0]{$M$}
\obj(100,10)[M1]{$M$}
\obj(150,10)[M2]{$M$}
%
%\obj(180,40)[eqivariance]{$e(p\cdot S) = e(p) \cdot \hat \ell(S)$}
%
\mor{LM}{M2}{$\>_{GL(m)}$}
\mor{P}{M1}{$\>_{SL(2,C)}$}
\mor{beP}{M0}{$\>_{SU(2)}$}
\mor{beP}{P}{$\io$}[\atleft,\solidarrow]
\mor{P}{LM}{$e$}[\atleft,\solidarrow]
\mor{M1}{M2}{}[\atright,\solidline]
\mor(100,13)(150,13){}[\atright,\solidline]
\mor{M0}{M1}{}[\atright,\solidline]
\mor(50,13)(100,13){}[\atright,\solidline]
}

We define an {\it $\SU(2)$-frame} a map $\ep \colon \Si\arr L(M)$, thus, for any spin frame $e \colon P\arr L(M)$, the map $e\circ \io \colon \Si\arr L(M)$ is a $\SU(2)$-frame (still on spacetime).
Locally, an $\SU(2)$-frame is still represented by an invertible matrix $e_I^\mu (x)\in \GL(4)$ (or its inverse $e^I_\mu(x)$).
What is characteristic of the $\SU(2)$-frame is the transformation laws, which are with respect to the automorphisms of $\Si$ instead of automorphisms of $P$.

\Note{
We have a group embedding $\Aut(\Si)\arr \Aut(P)$ and an element of $\Aut(\Si)$ is in the form
\eq{
\Cases{
x'^\mu=x'^\mu(x)\cr
U'= \psi(x)\cdot U
\qquad
\psi(x)\in \SU(2)
}
}
Let us denote by $\la \colon \SU(2)\arr \SO(3)\harr \GL(3)$ the covering map of the group $\Spin(3,0)=\SU(2)$, we have transformation laws
\eqLabel{
e'^\mu_I = J^\mu_\nu e^\nu_J \ell^J_I(\bar \psi)
\qquad\qquad
 \ell^J_I(\bar \psi) = \ell^J_I \circ i (\bar \psi) 
 =\(\Matrix{
 1	&0 \cr
 0	& \la(\bar \psi)
 }\)
}{SU2Automotphism}
Hence we split a tetrad $e_I^\mu$ into a vector $n = e_0$ and a triad $\ep_i := e_i$
}

When we then consider a $\Spin(3,1)$-connection $\om^{IJ}$, the situation is more complicated
since its horizontal spaces $H_p$ may or may not be tangent to the image $\io(\Si)\subset P$.
We need a way to project the horizontal subspaces onto the image to define a new connection $A^i$ on $\Si$, as well as some other field $k^i$ which carries the information we need to rebuild uniquely $\om^{IJ}$; see \cite{BI1}.

\ss{\ni\bf Definition:}
when we have a closed subgroup $H\subset G$, we say that $(G, H)$ is a {\it reductive pair} iff at the level of Lie algebras $\goth\subset \gotg$ the exact sequence of vector spaces
\eq{
0\arr \goth\arr \gotg \arr \gotm \arr 0
}
where we set $\gotm= \gotg/\goth$,
allows a {\it reductive splitting} $\Phi \colon \gotm \arr \gotg$, i.e.~the image $\Phi(\gotm)\subset \gotg$ is an invariant subspace with respect to the adjoint action of $G$ restricted to $H$, namely $T\Ad_G |_H \colon \Phi(\gotm) \arr \Phi(\gotm)$.
\ms

Notice that we are not requiring the reductive splitting to be a Lie algebra homomorphism, nor that $\gotm$ is a Lie subalgebra, which means that we are not requiring we have an exact sequence of groups, 
i.e.~$G$ does not need to split as a product of groups $G=H\times K$. 
Of course, if this is the case, as it was in the  Euclidean case $\Spin(4)= \SU(2)\times \SU(2)$ that Ashtekar originally considered in the selfdual-formalism (see \cite{Ash}, \cite{Rovelli1}), 
the pair $(G, H)$ is a reductive pair setting $\gotm=\gotk$ to be the Lie algebra of the subgroup $K$.

The same thing does not happen in the case $\Spin(3,1)=\SL(2,\C)$. 
We still have $\SU(2)\subset \SL(2, \C)$, but to define a complement one should have a sort of group generated by boosts, which unfortunately do not close to define a group.
More generally, at the level of algebras in any dimension we have the sequence
\Diagram{
\obj(0,10)[O1]{$0$}
\obj(50,10)[spinn]{$\spin(n,0)$}
\obj(120,10)[spinn1]{$\spin(n,1)$}
\obj(180,10)[m]{$\gotm$}
\obj(220,10)[O2]{$0$}
\mor{O1}{spinn}{}
\mor{spinn}{spinn1}{$i$}
\mor{spinn1}{m}{$p$}
\cmor((180,18)(176,25)(170,28)(150,30)(130,28)(122,25)(120,18)) \pdown(150,35){$\Phi$} [\atleft, \dashArrow]
\mor{m}{O2}{}
}
which sits in the corresponding Clifford algebra $\calC(n,1)$; see \cite{Lawson}, \cite{book2}.
The spin algebra $\spin(n,1)$ is spanned by the elements $e_{IJ}$, the spin algebra $\spin(n,0)$ is spanned by the elements $e_{ij}$,
$\gotm$ is spanned by $E_i:= p(e_{0i})$, which of course do not close to form a subalgebra.

We can define a splitting map $\Phi \colon \gotm\arr \gotg \colon E_i\mapsto e_{0i} + i\circ \be(E_i)$ for any linear map $\be \colon \gotm\arr \spin(n,0)\hlarr \spin(n,1)$, so that $p\circ i\circ \be=0$, by exactness.
We see immediately that $\Phi$ is a reductive splitting iff the map $\be \colon \gotm \arr  \spin(n,0) \colon E_i \mapsto  \be(E_i)=\be_i^{jk} e_{jk}$ is an intertwiner between $\gotm$, which supports the vector representation $\la$ of $\Spin(n,0)$, and $\spin(n,0)$, which supports the adjoint representation of $\Spin(n,0)$. 
Since both representations are irreducible, by Schur's Lemma this is possible only if $\dim(\gotm)=\dim(\spin(n,0))$, which is true only for $n=3$ (or $\be=0$).

In $n=3$, we hence have a whole family of reductive splittings $\be_i^{jk}= -\frac[1/2]\be \ep_i{}^{jk}$ parameterized by $\be$ which is called the {\it Immirzi parameter}.
Since the characteristic polynomial of $\be$ has at least a real root, then $\be\in \R$.
In all other dimensions, we have only one reductive splitting with $\be=0$.

 Let us remark that Holst parameter has a dynamical origin, while Immirzi parameter comes from kinematics. Setting {\it a priori} $\be=\ga$ is certainly possible, although rather suspect.

For any reductive splitting we have a different way of identifying $\spin(n,1)= \spin(n,0)\oplus \gotm$. In particular, for $n=3$ we have
\eq{
\frac[1/2]\om^{IJ} \si_{IJ} 
=  \om^{0i} \si_{0i}  + \frac[1/2]\om^{ij} \si_{ij} 
=  \om^{0i} \Phi\( E_{i}   \)   \oplus \frac[1/2] \( \om^{jk} + \be\ep_i{}^{jk}  \om^{0i}  \)  \si_{jk}   
}
We see that we can split a $\SL(2,\C)$-connection $\om^{IJ}_\mu$ as a pair $(A^i_\mu =\be \om^{0i}+  \frac[1/2]\ep^i{}_{jk} \om^{jk}  , k^i_\mu = \om^{0i}_\mu)$,
which is one-to-one map since it has inverse
\eq{
 \om^{jk}= \ep^{jk}{}_i \(A^i_\mu  -\be k^{i} \)
\qquad\qquad
  \om^{0i} = k^i
}

The map induced by $A = dx^\mu\otimes \(\del_\mu -A^i_\mu  \si_i\)$ defines in fact an $\SU(2)$-connection on the structure bundle $[\Si\arr M]$ which is called the {\it Barbero--Immirzi connection},
while $k = k^i_\mu dx^\mu \otimes \si_i$ is 1-form on $\Si$ valued in the Lie algebra $\su(2)$.
The expression of Barbero--Immirzi connection depends on a real parameter $\be$, the {\it Immirzi parameter}, since it relies on the reductive splitting.

The definition of Barbero--Immirzi connection is not only local but also global, in view of the globality of the reductive splitting. 
In fact, if one considers an automorphism $\phi\in \Aut(\Si)$, it induces an automorphism on $P$ (as we already used in (\ShowLabel{SU2Automotphism})),
which acts on $\om^{IJ}$ and hence on $(A^i, k^i)$. 

\Note{
One can check globality directly by using the special expression of (\ShowLabel{SU2Automotphism}). 
These are $\SU(2)$-gauge transformations, one can easily check that the transformation laws induced on 
$(A^i, k^i)$ are in fact
\eq{
A'^i_\mu = \barJ^\nu_ \mu\( \la^i_j  A^j_\nu  -\frac[1/2]\ep^{il}{}_j\la^j_m \di_\nu \bar \la^m_l\)
\qquad
k'^i_\mu = \la^i_j(\psi) k^j_\nu \barJ_\mu^\nu 
}
which are what is expected from an $\SU(2)$-connection and a $1$-form valued in $\su(2)$.
}

Since $A$ is a global connection on $\Si$, it defines its {\it curvature 2-form} $F^k = \frac[1/2] F^k_{\mu\nu}  \> dx^\mu \land dx^\nu$ where we set
\eq{
F^i_{\mu\nu} = d_\mu A^k_\nu-  d_\nu A^k_\mu - \ep^k{}_{ij} A^i_\mu A^j_\nu
}
which transforms in the adjoint representation of $\su(2)$.
This (as well as the invariance of holonomy we shall use later on) is obtained only in view of the globality of $A$ as an $\SU(2)$-connection. 
Globality of $A$ is crucial, even when $\Si$ is trivial; not (only) for geometric reasons, but (and more importantly) for ensuring the correct transformation laws of objects
which are important from a physical perspective in the first place.

We can now write the Holst Lagrangian in terms of the new fields $(e^I_\mu, A^i_\mu, k^i_\mu)$ as
\eqs{
L_H=& \frac[1/\ka ]      F^i \land  L_i   + \frac[1/\ka ]  \na k^{i} \land \( K_i-\be L_i\)     - \frac[1/2\ka] \ep_{ijk}  k^{i}\land   k^{j}\land \(  \(\be^2-1\)  L^k -2\be K^k \)  + \cr
& +   \frac[\La/3 \ka] \frac[\ga^2 /1-\ga^2]  K^k \land L_k 
}
where we set $K_k = \frac[1/2] \ep_{kij}  e^i\land e^j  - \frac[1/\ga] e^0\land e_k$ and $L_k =   e^0\land e_k  + \frac[1/2\ga] \ep_k{}^{ij} e_i\land e_j $,
for the {\it boost} and {\it rotational parts} of the $B_{IJ}$ field, respectively; see \cite{Rovelli2}.
Here the covariant derivative $\na$ is the covariant derivative with respect to the $\SU(2)$-connection $A^i$. 
In particular, we have $ \na k^k =  d k^{i}   - \ep^{i}{}_{jk}  A^j  \land k^{k}$.

\Note{
Let us stress that since this is a Lagrangian for the fundamental fields $(e^I_\mu, A^i_\mu, k^i_\mu)$, it must be varied with respect to its fundamental fields, not with respect to $(K^i_{\mu\nu}, L^i_{\mu\nu}, A^i_\mu, k^i_\mu)$.
Variations with respect to $(K^i_{\mu\nu}, L^i_{\mu\nu}, A^i_\mu, k^i_\mu)$ are, in fact, not independent.

To find the Lagrangian in the Barbero--Immirzi formulation, one can use the identities
\eq{
R^{0i}=  \na k^{i}     + \be \ep^{i}{}_{jk}  k^{j}\land   k^{k}
\qquad\qquad
\frac[1/2]\ep^k{}_{ij}R^{ij} = F^k -\be\na k^{k}   -  \frac[\be^2-1/2] \ep^{k}{}_{ij}  k^{i}\land  k^{j}
}
which relate the curvature of $\om^{IJ}$ to the curvature of $A^i$. 
}

Field equations in the Barbero-Immirzi formalism read as
\eqLabel{
\Cases{
	\na L_k =     \frac[\be -\ga/\ga ]   k^{i} \land e_i  \land e_k  + \frac[\ga\be+1 /\ga^2+1 ]   \ep_{kij}  k^{i}  \land  (K^j - \ga L^j)  	 \cr
	\na K_k  =\frac[\be\ga+1 /\ga ]    k^{i} \land e_i\land e_k  + \frac[ \ga -\be/ \ga^2+1 ]  \ep_{kij}  k^{i} \land   (K^j - \ga L^j)    \cr
  F^k \land  e_k   -  \frac[1 +\be\ga/\ga]    \na k^{k} \land  e_k 
+  \frac[\ga-2\be-\ga\be^2/2\ga]  \ep^k{}_{ij} k^{i}\land   k^{j} \land  e_k   
+ \frac[\La/6]  \ep_{ijk } e^i\land e^j\land e^k
           =0\cr
  \( \ga F^h -  (1 +\be\ga)   \na k^{h} +  \frac[\ga-2\be-\ga\be^2/2]   \ep^h{}_{ij}  k^{i}\land   k^{j}  - \frac[\La/2]  \ep^h{}_{ij}  e^i\land e^j  \) \land  e^0  +\cr
\qquad+ \(  \ep^h{}_{kj}  F^k\land e^j   +  (\ga-\be)   \ep^h{}_{kj} \na k^{k}\land  e^j    + \( \be^2-1-2\be\ga \)  k^{h}\land    k_{l}\land   e^l  \) =0
}
}{BIFieldEquations}

The first two field equations can be recast as
\eq{
\ep^k{}_{ij} \hat \na  e^i\land e^j =0 
\qquad\qquad
 \hat \na e^0 \land e^k  =   \hat \na e^k \land e^0   
}

These are definitely not simple or beautiful. Keep the sensation in mind because it will call for a miracle when we shall see how simple they get after the (rather horrific) decomposition along a foliation; \cite{ADM}, \cite{NostroADM}, \cite{Gar1}. 
In particular we shall see that, without assuming anything about the Holst and the Immirzi parameters, the Holst parameter disappears from constraint equations and they depend on the Immirzi parameter $\be$ only.
Moreover, we are able to completely solve algebraic field equations to completely determine $k^i$ as a function of the frame, in the bulk actually.
Finally, we shall see that on the foliation we will have $K_i=\be L_i$, which will be important later in the quantum theory; see \cite{Rovelli2}.
%This relation is obtained in the bulk as well if we set $\be=\ga$.
In any event, these field equations are just obtained from the Holst ones by a field transformation to define $(A, k)$, hence they are dynamically equivalent to Holst equations and standard GR.

\section{Conclusions and Perspectives}

Although field equations in the Barbero--Immirzi formalism on spacetime are not particularly appealing, this is a good place to stop the first lecture.
All the topological arguments to ensure existence of global structures have been discussed here.

The relevant structures we introduced here are:

\Itemize{
\Item[-] the spacetime $M$ has to be a {\it spin manifold}, so that it allows spin structures, which are required both for the existence of spin frames and for maintaining the possibility of having global Dirac equations.
This guarantees the existence of a structure bundle $P$ as well as a global spin frame $e \colon P\arr L(M)$, which we use as fundamental fields.
\Item[-] the $\SU(2)$-reduction $(\Si, \io)$ of $P$ from the $\Spin(3,1)$ group to $\Spin(3,0)$. 
For a spin manifold, this comes for free, with no topological obstruction; see \cite{Orizzonte}. It is used here to define the structure bundle $\Si$ on which we define the Barbero--Immirzi fields $(A, k)$.
Later on, this will also be used to adapt frames to the foliation. LQG is a $\SU(2)$-gauge theory hence this structure is particularly important.
\Item[-] the reductive splittings which are used in $n=3$ to project the spin connection $\om^{IJ}$ to $\SU(2)$ fields $(A, k)$.
This is a weaker structure than the group splitting $\Spin(4,0)=\SU(2)\times \SU(2)$ which was originally used for selfdual formalism.
Reductive splittings are needed to deal with the group $\Spin(3,1)$ which does not split, without resorting to complexification.
}

\Note{
It has been argued that Barbero-Immirzi connection cannot be defined on spacetime.
The argument goes like showing that one can fix a particular spacetime and a foliation, define the Barbero-Immirzi connection $A$ on a leaf and then compare the holonomy along a path $\ga$ on a leaf of the two connections $A$ and $\om$; see \cite{Samuel}, \cite{HolonomyClassified}.
One can show that in some cases the result is different and that shows that the Barbero-Immirzi connection $A^i$ cannot be a restriction of the spin connection $\om^{IJ}$.
In fact it is not, neither for us it is. The connection $\om^{IJ}$ is not tangent to $\Si$, it has to be projected and splitted as $(A^i, k^i)$ to become tangent to $\Si$.
The connection $\om^{IJ}$ is not {\it restricted} to $\Si$, it is  {\it projected} on it.
Accordingly, there is no need for the holonomies of $\om^{IJ}$ and $A^i$ to be the same (and in fact they are not).
The holonomy of $A^i$ is not the original holonomy of $\om^{IJ}$, it just encodes it together with $k^i$.
Nevertheless, we could define a global $\SU(2)$-connection $A^i$ on spacetime, which can be then restricted to the leaves of a foliation to define the standard Barbero-Immirzi on space.
}

Of course, one can assume structures and ignore these topological arguments.
The physical important fact is that fundamental fields transform as they should to be global (as they do and one can check it directly).
This is important because in many instances we shall consider structures (e.g., curvature, holonomy, tensors) just because they transform in a given way and they would not if the fundamental fields did not transform as they do.
Here topological arguments are mainly a motivation for definitions which otherwise would come out of the blue.

Notice that $K$ and $L$ are 2-forms induced by the spin frame, thus given a frame the first two equations are algebraic in the field $k^i$, while the other equations involve the curvature $F^i$ of the Barbero--Immirzi connection.

In the next lecture we shall give a unifying framework to discuss Cauchy problems and pre-quantum equations. Although the discussion applies to quite a general field theory (including Maxwell equations), we will apply it to Barbero--Immirzi formulation of standard GR; see \cite{Gar1}, \cite{Gar2}.

Mathematically speaking, field equations for a generic Lagrangian are quasi-linear, although usually they are not elliptic nor hyperbolic. Then one can split the field equations (as well as the fields) in two parts, one which is hyperbolic and one which are constraints on initial conditions, which account for over-determination of Einstein (or Maxwell) equations. These initial conditions produce the Cauchy data for the Cauchy problem, which is defined for the hyperbolic part of the equations and fields.
Usually, this splitting is done in Hamiltonian formalism as the canonical analysis of the theory. We shall do it first in Lagrangian formalism. It is not a property of a particular formalism for field theory, it is a property of field equations
whatever formalism one uses to write them in.

Depending on the interest, one can solve constraints for allowed initial conditions, find a solution of Cauchy problem and rebuild a covariant solution of the original covariant field equations. 
This is a classical attitude, which is more or less a starting point for classical numerical gravity; see \cite{Gourgoulhon}.

Alternatively, one can realize that determining bulk fields is a classical goal that is unrealistic in a quantum context. 
In mechanics, it would correspond to trace the trajectory of particles, which, we know, does not make sense in quantum mechanics. 
A quantum attitude is to forget what happens in the bulk and focus only on the boundary to assign a probability amplitude to the propagation of boundary data. 
For this reason, one should quantize constraint equations, then possibly average for a classical initial condition (e.g.\ by using some coherent state formulation) and then solve a classical Cauchy problem to rebuild the covariant field. 
This route is the pre-quantum formalism and it is followed by LQG, which in fact quantizes constraint equations of the Barbero--Immirzi model.
In other words, the splitting of field equations will provide us with an initial step both for classical numerical solutions and for quantization.

\section*{Appendix. Structures on spacetimes and background free models.}

This is almost trivial but it is worth saying explicitly.
There are two things we call {\it GR}, which are different and must be kept distinct.

When we describe the motion of planets around a star, that of a star in the galaxy, the evolution of density perturbations during the expansion of the universe,
or the propagation of gravitational waves, we describe the spacetime as a manifold $M$ {\it with} a fixed Lorentzian metric $g$ on top of it.
In this context, spacetime is a Riemannian manifold $(M, g)$.
Mathematically speaking, studying the properties of $(M, g)$  is what one does in {\it differential geometry}. 
Also in cosmology, when we fix the cosmological principle, we obtain an ansatz for the metric to be used and we try to adjust the metric to observations, which is pure and clear differential geometry.

On the contrary, when we discuss (quantum or classical) field equations for gravity, e.g.~by fixing a variational principle, spacetime is a {\it bare manifold} $M$, which {\it allows} Lorentzian metrics, but we do not fix any metric (or any other structure) on it.
We write a variational principle for {\it any} metric, we obtain field equations, and we find metrics as solutions of these field equations.
Here most of the work is done with no metric fixed on $M$, the metric is the result of the process, but spacetime is a bare manifold. 
In a quantum setting, one does not even quantize a spacetime metric at all, we shall see that in LQG we quantize a conjugated pair made of an $\SU(2)$-connection and a (densitized) triad on space from which the Lorentzian covariant metric will eventually emerge.
In mathematics, studying the property of a bare manifold $M$  is called {\it differential topology}. 

Let us remark that one of Riemann's motivations for introducing abstract manifolds was exactly to point out that a manifold exists first as a bare manifold, with no geometric structure fixed on it.
Then, eventually, it can be embedded in an environmental space so that it inherits an induced metric from the metric in the environmental space in which it is embedded into.
That is the novelty introduced by Riemann over Gauss and his generation, who studied {\it surfaces} which are defined as {\it embedded manifolds}. 

In the context of GR, manifolds (and spacetimes) are {\it bare} manifolds. They become Riemannian manifolds when one fixes a metric on them.

\Note{
As a consequence, when discussing Einstein equations, one should not say that leaves of an ADM foliation are {\it spacelike submanifolds}, since there is no metric to be spacelike for.
On the contrary, one fixes a foliation, solves the equations, defines a covariant metric and proves that the leaves are {\it by construction} spacelike with respect to the resulting metric.
As a matter of fact, the foliation being spacelike is a property of field equations, not a property of the foliation.
}
 
Let us point out that one can do a lot of mathematics on bare manifolds. 
One can define tangent vectors, tensor fields, discuss topological obstructions to the existence of tensor fields with given properties, define variational principles and global PDEs,
discuss Cauchy problems (see \cite{Choquet}, \cite{Gar1}, \cite{Gar2}), flows of vector fields, and so on.
Let us also stress that a {\it background free theory} is {\it precisely} a theory written on a bare manifold, something which does not depend on any structure fixed (by us) on the manifold. 
That is one of the (not so) hidden assumptions of GR: all the structures on spacetime have to be determined by equations, all structures are dynamical.
This is an important axiom since it is not trivial to write a variational principle with a given set of fields without introducing other fixed structures. That issue sits at the core of the discussion between Einstein and Kretschmann (1917) 
which dates back to the origin of GR and one of the times in which Einstein was too fast in acknowledging to be wrong; see \cite{Kretschmann}, \cite{Giulini}, \cite{HoleArgument}.

\Note{
The issue is not completely settled, however. 
As a matter of fact, one does have structures (fields) fixed on a bare manifold.
For example the  Kronecker delta is a $(1,1)$-tensor field, the Levi-Civita symbols define an $(m, 0)$ and a $(0, m)$-tensor density.
These are {\it canonical} structures and nobody has ever argued they need to be varied or need to be determined by equations.

The problem is that there is no clear-cut definition of canonical structures, which are exceptions to the rule above. 
One example is when one fixes a signature and defines $\eta_{IJ}$ which determines what is an orthonormal frame. 
These $\eta_{IJ}$ live in the algebra of frames, not on the manifold and they are considered to be a canonical structure.
This is a particularly beautiful example, since it has not to be confused with the metric (with coefficients $\eta_{\mu\nu}$, written with world indices, not with frame indices) in Minkowski spacetime, which being a metric on spacetime is a structure and fixing it to do special relativity (SR) is the reason why SR is not a relativistic theory.

On a (gauge-)natural bundle, one can define a structure to be canonical if it is invariant with respect to the action of the group of diffeomorphisms $\Diff(M)$ (or generalized gauge transformations $\Aut(P)$), which in fact acts on (gauge-)natural bundles.
This covers the examples of canonical structures given above. In a more general setting, one needs to discuss what is meant by canonical structure and, consequently, by background free.
} 
 
Anyway, as a matter of fact, a relativistic theory (and GR in particular) is a field theory on a bare manifold, in which all (non-canonical) structures are dynamically determined by field equations. 
When one fixes which fields are involved (as well as to which order they enter in the action) that does impose strong constraints on the allowed dynamics (Utiyama theorems; see \cite{Janiska}, \cite{Kolar}, \cite{book2}),
which eventually contradicts what Kretschmann said.

\Note{
Originally Einstein was presenting GR as a theory based on general covariance, saying that general covariance was the core of the new theory.
At one of his first lectures, Kretschmann argued that, mathematically speaking {\it any} equation can be written in general covariant form at the price of adding more fields and possibly more equations.
That is obviously true, but the argument was never formulated properly, being precise on what one had to understand by a {\it field theory} and what did it mean to add fields to it.
Instead Einstein, based on few examples provided by Kretschmann, conceded that he was right and started introducing the theory as based in {\it equivalence principle} which soon became recognised as the physical core of GR.

Later on, also discussing gauge theories, we learned clearly that discussing the issue starting from a variantional principle {\it and fixing} the set of fields involved in the theory, general covariance (as well as gauge covariance)
as well as requiring the theory to be background free, actually imposes strong constraints on allowed dynamics. The issue was actually coded into a family of results called the {\it Utiyama theorem}.

Not only one did not need to abandon general covariance as the core of a relativistic theory, rather the other way around, one can show that a combination of general covariance and background freedom actually implies equivalence principle. 
As a matter of fact, one can even prove that geodesics equations for freely falling material points (which are in fact a form of weak equivalence principle) are in fact the simplest equations one can write on a bare manifold, based only on covariance requirements,
showing in some sense that (weak) equivalence principle follows from covariance requirements; see \cite{Geodesics}, \cite{book2}, \cite{EPS}.
Moreover, one has one of such equations for any {\it projective class of connections} which then arise as a  natural candidate of a (part of the) description of gravitational field.
}

Finally, let us mention that, in differential topology, one does not really work on a bare manifold. 
There are (global) diffeomorphisms acting on bare manifolds which are required to be symmetries in gravitational theories.
As a result, one defines an equivalence relation among bare manifolds (two bare manifolds are equivalent if they are (globally) diffeomorphic) and takes the quotient.
Spacetime is not a bare manifold, it is a bare manifold {\it up to diffeomorphisms}.
A {\it geometry} on spacetime is not a metric on $M$, it is a whole class $[(M, g)]=\{(\phi(M), \phi_\ast g)\}$ of isometric Riemannian manifolds.

Einstein equations are not equations for a metric on a bare manifold,  they are {\it covariant} equations for metrics on a bare manifold, which exactly means that they are compatible with the quotient, hence they induce ``equations'' for classes on the quotient.
On the quotient they are equations for the gravitational field which is represented by classes $[(M, g)]$.
The physical equations are on the quotient, they are defined just as far as the equations on the manifold are covariant.
Of course, if the quotient were a manifold itself, one could write PDEs directly on the quotient. However, this is not the case and, with the current technology, the only consistent way of writing equations for classes of bare manifolds is to write 
covariant PDEs on bare manifolds which represent points in the quotient space.

By the way, that is why one says {\it points on spacetime are not observables}, meaning precisely that the (local) coordinate functions $x^\mu(p)$ are not diffeomorphic invariant, they are a characteristic of the bare manifold $M$, 
not of the class $[M]$, which is where the physical theory is defined.
A precise language has been developed in differential topology to discuss these kind of issues. 
Ignoring it and adding instead a new imprecise one to go along our physical intuition (which in this case goes back to Newton and the need to rely on an absolute  space or pretend to) is not a good idea in this case.

Moreover, the mathematical language adapts perfectly to physics. Bare manifolds are covered by local observers (their charts). Ignoring specific representatives means exactly looking for properties which are independent of observers, absolute properties which are, by definition, the real physics. Real physics is in the quotient, on bare manifolds up to diffeomorphisms. The gravitational field is {\it geometry} which is in the quotient.
It is not a metric, it is a class of equivalent Lorentzian manifolds.
Unfortunately, we are not able to give dynamics directly on the quotient as we are not able to describe physics (meaning real world observations) without using observers.
Bare manifolds encode what observers measure, they are relative to the observations.
These two worlds are connected, precisely, by diffeomorphisms, which encode changes of observers. Physical information is in the relations among observes, not in the observers themselves.

\section*{Acknowledgements}
We also acknowledge the contribution of INFN (Iniziativa Specifica QGSKY and Iniziativa Specifica Euclid), the local research project {\it  Metodi Geometrici in Fisica Matematica e Applicazioni (2023)} of Dipartimento di Matematica of University of Torino (Italy). This paper is also supported by INdAM-GNFM.

We are grateful to C.Rovelli and S.Speziale for comments and discussions.

L. Fatibene would like to acknowledge the hospitality and financial support of the Department of Applied Mathematics, University of Waterloo where part of this research was done.


\medskip


\begin{thebibliography}{99}

\bibitem{Ash}{A.Ashtekar,
{\it New Perspectives in Canonical Gravity},
Bibliopolis, Naples, (1988)
}

\bibitem{Rovelli1}{C.Rovelli, 
{\it Quantum Gravity}, 
Cambridge University Press, (2004) 
}

\bibitem{Rovelli2}{C.Rovelli, F.Vidotto, 
{\it An elementary introduction to Quantum Gravity and Spinfoam Theory},
Cambridge University Press, (2014)
}

\bibitem{Samuel}{J.Samuel,
{\it Is Barbero's Hamiltonian formulation a gauge theory of Lorentzian gravity?},
Classical and Quantum Gravity, {\bf 17},  (2000) ; % L141-L148
{\tt arXiv:gr-qc/0005095}
}

\bibitem{HolonomyClassified}{M. Berger, 
{\it Sur les groupes d’holonomie des vari\'et\'es \'a connexion affine et des vari\'et\'es Riemanniennes}, 
Bull. Soc. Math. France 83, (1955) %279-330
}

\bibitem{Steenrod}{N.Steenrod,
{\it The Topology of Fibre Bundles}, 
PMS {\bf 14},
Princeton University Press, (1951);\goodbreak
{\tt https://www.jstor.org/stable/j.ctt1bpm9t5}
} 

\bibitem{KN}{S.Kobayashi,  K.Nomizu,  
{\it Foundations of Differential Geometry}. 
Wiley Classics Library. Wiley (1963)
}

\bibitem{book1}{L.Fatibene, M.Francaviglia,
{\it Natural and Gauge-Natural theories},
Kluwer, Dordrecht, (2003)
}

\bibitem{book2}{L.Fatibene,
{\it Relativistic theories, gravitational theories and General Relativity},
in preparation, draft version 1.0.2.
{\tt http://www.fatibene.org/book.html}
}

\bibitem{Pathway}{C.Rovelli, 
{\it Halfway through the woods}, in: {\it The Cosmos of Science}, 
J Earman and JD Norton ed., University of Pittsburgh Press and Universit\"ats Verlag-Konstanz (1997)
}

\bibitem{Giulini}{D.Giulini,
{\it Some remarks on the notions of general covariance and background independence},
Lect. Notes Phys. 721 (2007); % 105-120;
{\tt arXiv:gr-qc/0603087}
}

\bibitem{HolstOriginal}{S. Holst, 
{\it Barbero’s Hamiltonian derived from a generalized Hilbert-Palatini action},
Phys. Rev. D 53 (1996); % 5966;
{\tt arXiv:gr-qc/9511026}
}

\bibitem{NostroRov1}{L. Fatibene, M.Francaviglia, C.Rovelli, 
{\it Lagrangian Formulation of Ashtekar-Barbero-Immirzi Gravity}, 
CQG 24 (2007); % 4207-4217; 
{\tt gr-qc/0706.1899}
}


\bibitem{NostroRov2}{L. Fatibene, M.Francaviglia, C.Rovelli, 
{\it On a Covariant Formulation of the Barberi-Immirzi Connection},
CQG 24 (2007); % 3055-3066;
{\tt gr-qc/0702134v1}
}

\bibitem{NostroHolst}{
L. Fatibene, M. Ferraris, M. Francaviglia, G. Pacchiella, 
{\it Entropy of Self-Gravitating Systems from Holst's Lagrangian},
Int. J. Geom. Methods Mod. Phys. {\bf 6}(2) (2009) %361-365
}

\bibitem{ADM}{R.Arnowitt, S.Deser, C.W.Misner, in: 
{\it Gravitation: An Introduction to Current Research}, 
L. Witten ed. Wyley,  227, New York, (1962)
}

\bibitem{NostroADM}{L.Fatibene, M.Ferraris, M.Francaviglia, L.Lusanna, 
{\it ADM Pseudotensors, Conserved Quantities and Covariant Conservation Laws in General Relativity}, 
Annals of Physics, {\bf 327}(6), (2012); %, Pages 1593?1616;
{\tt arXiv:1007.4071 [gr-qc]}
}

\bibitem{Thiemann2}{T. Thiemann, 
{\it Modern Canonical Quantum General Relativity}, 
(Cambridge University Press, New York, 2017)
}

\bibitem{Lawson}{H.B.Lawson Jr., M.L.Michelsohn, 
{\it Spin Geometry},
Princeton University Press, New Jersey, (1989)
}

\bibitem{Jadwisin}{L. Fatibene, M. Francaviglia,
{\it Deformations of spin structures and gravity},
in: Gauge theories of gravitation (Jadwisin, 1997). 
Acta Phys. Polon. B {\bf 29}(4) (1998), %915-928.
}

\bibitem{Spinors}{
L. Fatibene, M. Ferraris, M Francaviglia, M. Godina,
{\it Gauge formalism for general relativity and fermionic matter},
GRG {\bf 30}(9) (1998), %1371?1389.
}

\bibitem{Noris}{R.Noris, L.Fatibene,
{\it Spin frame transformations and Dirac equations},
Int. J. Geom. Methods Mod. Phys.  {\bf 19}(1),  (2022) % 2250004
}

\bibitem{Orizzonte}{A.Orizzonte, L.Fatibene,
{\it Barbero--Immirzi connections and how to build them}
Journal of Geometry and Physics {\bf 167} (2021)%, 104276
}

\bibitem{Kolar}{I.Kol\'a\u r, J.Slov\'ak, P.W.Michor,
{\it Natural Operations in Differential Geometry},
Springer-Verlag Berlin Heidelberg (1993)
}

\bibitem{LN6}{L.Fatibene,  A.Orizzonte,
{\it Lecture Notes in Loop Quantum Gravity. LN6: $SU(2)$ representations and intertwiners},
(to appear)}

\bibitem{NostroAsh}{L. Fatibene, M. Francaviglia,
{\it Spin Structures on Manifolds and Ashtekar Variables}, 
Int. J. Geom. Methods Mod. Phys. 2 (2005)%, no. 2,147-157
}

\bibitem{BI1}{
L.Fatibene, M.Ferraris, M.Francaviglia,
{\it Inducing Barbero-Immirzi connections along SU(2)-reductions of bundles on spacetime},
Phys. Rev. D {\bf 84},  (2011); %064035
{\tt arXiv:1011.2041}
}

\bibitem{Gar1}{L.Fatibene, S. Garruto,
{\it The Cauchy problem in General Relativity: An algebraic characterization},
CQG {\bf 32}(23) (2015); % , 235010;  
{\tt arXiv:1507.00476}
}


\bibitem{Gar2}{Lorenzo Fatibene, Simon Garruto,
{\it Principal Symbol of Euler-Lagrange Operators},
CQG {\bf 33}(14), (2016);\goodbreak
{\tt arXiv:1603.04732}
}

\bibitem{Gourgoulhon}{\'Eric Gourgoulhon,
{\it 3+1 Formalism and Numerical Relativity},
General Relativity Trimester, Institut Henri Poincar\'e
}


\bibitem{Choquet}{Y.Choquet-Bruhat, R.Geroch,
{\it Global aspects of the Cauchy problem in general relativity}, 
Comm. Math. Phys. {\bf 14},  (1969) %329–335
} 


\bibitem{Kretschmann}{E.Kretschmann,
{\it \"Uber den physikalischen Sinn der Relativit\"atspostulate, A.Einsteins neue und seine urspr\"ungliche Relativit\"atstheorie}, 
Annalen der Physik, {\bf 53}, (1917), %575Ð614
}

\bibitem{HoleArgument}{
L. Fatibene, M. Ferraris, G. Magnano,
{\it Constraining the Physical State by Symmetries},
Annals of Physics 378, (2017); % 47-58; 
{\tt arXiv:1605.03888}
}


\bibitem{Janiska}{J.Jany\v ska,
{\it Utiyama's reduction method and infinitesimal symmetries of invariant Lagrangians},
Symmetry and Perturbation Theory,  (2007) % 255-256
}


\bibitem{Geodesics}{L.Fatibene, M.Francaviglia,  G.Magnano,
{\it On a characterization of geodesic trajectories and gravitational motions},
Int. J. Geom. Meth. Mod. Phys. {\bf 09}(5),  (2012)	%1220007
}


\bibitem{EPS}{J.Ehlers, F.A.E.Pirani, A.Schild, 
{\it The Geometry of Free Fall and Light Propagation}, 
in: {\it General Relativity}, 
ed. L.ORaifeartaigh (Clarendon, Oxford, 1972).
}


\end{thebibliography}
\end{document}